\documentclass[
  journal=pasa,
  manuscript=article-type,
  year=2020,
  volume=37,
]{cup-journal}

\usepackage{aas-macros} % PASA wants this

\makeatletter
\renewcommand\@biblabel[1]{[#1]}% plain brackets, no \textbf
\makeatother
\usepackage{rotating}
\usepackage{amsmath}
\usepackage{longtable}
\usepackage{amssymb}
\usepackage{hyperref}
\usepackage[nopatch]{microtype}
\usepackage{booktabs}
\usepackage{lscape}
\usepackage{longtable}
\usepackage{subfig}
\usepackage{supertabular,booktabs}
\usepackage{gensymb}
\usepackage{float}
\usepackage{lscape} 
\usepackage{aas-macros}

\title{The Secret Lives of Open Clusters: a Multiwavelength Examination of Three Open Clusters}

\author{Kristen C. Dage}
\affiliation{International Centre for Radio Astronomy Research $--$ Curtin University, GPO Box U1987, Perth, WA 6845, Australia}

\author{Emily L. Hunt}
\affiliation{Department of Astrophysics, University of Vienna, Türkenschanzstrasse 17, 1180 Wien, Austria}
\affiliation{Max-Planck-Institut für Astronomie, Königstuhl 17, 69117 Heidelberg, Germany}

\author{Jasmine Anderson-Baldwin}
\affiliation{Centre for Astrophysics and Supercomputing, Swinburne University of Technology, Hawthorn, VIC 3122, Australia}
\affiliation{ARC Centre of Excellence for Gravitational Wave Discovery (OzGrav), Hawthorn, VIC 3122, Australia}

\author{Evangelia Tremou}
\affiliation{National Radio Astronomy Observatory, Socorro, NM 87801, USA}

\author{Khushboo K. Rao}
\affiliation{Institute of Astronomy, National Central University, Taiwan}

\author{Kwangmin Oh}
\affiliation{Center for Data Intensive and Time Domain Astronomy, Department of Physics and Astronomy, Michigan State University, East Lansing, MI 48824, USA }

\author{Malu Sudha}
\affiliation{Department of Physics \& Astronomy, Wayne State University, 666 W. Hancock St, Detroit, MI 48201, USA}

\author{Jarrod Hurley}
\affiliation{Centre for Astrophysics and Supercomputing, Swinburne University of Technology, Hawthorn, VIC 3122, Australia}
\affiliation{ARC Centre of Excellence for Gravitational Wave Discovery (OzGrav), Hawthorn, VIC 3122, Australia}

\author{Robert D. Mathieu}
\affiliation{Department of Astronomy, University of Wisconsin - Madison, Madison WI, USA}
\author{Aarya Patil}
\affiliation{Max-Planck-Institut f¨ur Astronomie, K¨onigstuhl 17, D-69117 Heidelberg, Germany}

\author{Richard M. Plotkin}
\affiliation{Department of Physics, University of Nevada, Reno, NV 89557, USA}
\affiliation{Nevada Center for Astrophysics, University of Nevada, Las Vegas, NV 89154, USA}

\author{Andrew M. Hopkins}
\affiliation{School of Mathematical and Physical Sciences, 12 Wally’s Walk, Macquarie University, NSW 2109, Australia}

\author{Jacco Th. van Loon}
\affiliation{Lennard-Jones Laboratories, Keele University, ST5 5BG, UK}

\author{Jayde Willingham}
\affiliation{School of Mathematical and Physical Sciences, 12 Wally’s Walk, Macquarie University, NSW 2109, Australia}

 %% First letter not capped

\begin{document}

\begin{abstract}
Star clusters are well known for their dynamical interactions, an outcome of their high stellar densities; in this paper we use multiwavelength observations to search for the unique outcomes of these interactions in three nearby Galactic open clusters: IC 2602 (30  Myr), NGC 2632 (750 Myr) and M67 (4 Gyr). We compared X-ray observations from all-sky surveys like eROSITA, plus archival observations from \textit{Chandra} X-ray Observatory, survey radio observations from ASKAP's Evolutionary Map of the Universe survey plus archival VLA observations, in conjunction with new cluster catalogs with Gaia. From X-ray, we found 77 X-ray sources likely associated with IC 2602, 31 X-ray sources in NGC 2632, and 31 near M67's central regions. We were further able to classify these X-ray sources based on their optical variability and any radio emission. Three IC 2602 X-ray sources had radio counterparts, which are likely all chromospherically active binary stars. We also identified luminous radio and X-ray variability from a spectroscopic triple system in M67, WOCS 3012/S1077, which is either consistent with a quiescent black hole binary, or due to an active binary stellar system. A recent population study of optical variables by Anderson \& Hunt 2025 shows that the population of optical variables in open clusters clearly changes over \textbf{cluster age}; this pilot study gives evidence that the X-ray population also changes with time, and demonstrates the need for a broader multiwavelength study of Galactic open clusters. 

\end{abstract}

\section{Introduction}  
Open clusters (OCs) are densely bound collections of stars, born out of the collapse of large clouds of gas. They can range in age from a few tens of Myr to a few Gyr, and are single stellar populations \citep[SSPs][]{PortegiesZwartMcMillan_2010,Cantat-GaudinCasamiquela_2024}. Due to this, they are useful as tracers of Galactic chemical trends \citep{2020AJ....159..199D}, as well as understanding simple stellar evolution in a dynamical environment \citep{hurley05, Hunt24}. Because of their distribution in the Milky Way and proximity to Earth, their individual stars can be resolved and well-studied. 

Data from ESA's Gaia satellite \citep{Gaia2016} has completely revolutionized the census of OCs in just a few short years \citep{Cantat-GaudinCasamiquela_2024}. Since the release of Gaia DR2 \citep{Gaia18}, thousands of new OCs have been discovered \citep[e.g.][]{Castro-GinardJordi_2018,LiuPang_2019,HuntReffert_2023}, over a thousand OCs from before Gaia have been ruled out as asterisms \citep{Cantat-GaudinAnders_2020,HuntReffert_2023}, and the general quality of the OC census has improved, both in terms of improved membership determination \citep{Cantat-GaudinJordi_2018,DiasMonteiro_2021,HuntReffert_2023,PerrenPera_2023} and improved determination of parameters such as cluster distance, age, and mass \citep{Cantat-GaudinAnders_2020,HuntReffert_2023,CavalloSpina_2024}. There has hence never been a better time to do science with OCs in the Milky Way.

With the available wealth of X-ray and radio surveys, it is now possible to delve into the contents of OCs by searching for their multiwavelength counterparts. For example,  \cite{belloni98, vdb04} and \cite{mooley15}  used X-ray observations from \textit{Chandra} and XMM-Newton to identify X-ray counterparts to individual cluster members of the old open cluster M67. Radio emission is also helpful to classify X-ray sources \citep[e.g.,][among others]{2024PASA...41...84D, Paduano24}

Two major sources of uncertainty for these kinds of studies are 1) determining whether a given X-ray point source has an optical counterpart (which is dependent on positional accuracy), and 2) whether that optical counterpart is a cluster member. While the association of multiwavelength sources to OC members is challenging, it is useful to find unique sources to benchmark SSP evolution in a dense field. 

In old open clusters (older than a few Gyr), most single stars have spun down due to magnetic braking and are consequently faint in X-ray \citep[e.g.][]{Caillault1996}. In clusters around this age, bright X-ray sources are instead dominated by binaries. In binaries with relatively short separations (or relatively large radii), tidal interactions can force the stars into a higher rotation rate, leading to higher X-ray emission \citep[e.g.][]{belloni98}. In other cases, accretion onto compact objects within a binary also produces luminous X-ray sources. Multiwavelength observations, particularly in the X-ray, allow us to probe the populations of these close binary systems. Since binary populations are drivers of cluster evolution \citep{Hut92}, this in turn gives us greater understanding of the dynamical evolution of clusters.

Beyond the influence that dynamics provide to enhance BH formation in clusters, from the observational standpoint, it is natural to associate the origins of many black holes with star clusters, including new evidence in the form of Gaia BH3 \citep{2024A&A...686L...2G}, which is found in the stellar stream of ED-2, a disrupted low mass cluster \citep{2024A&A...687L...3B}. 
In this paper, we use archival and survey observations from X-ray and radio facilities to search for multiwavelength counterparts to individual OC stars, as identified by \cite{Hunt24}. We specifically target nearby OCs from three different ages, and compare their unique multiwavelength contents, with the aim to use this information to provide a benchmark for the contents of open clusters of different ages and masses to improve future iterations of N-body simulations of open clusters.

\section{Data and Analysis}
For our cluster sample, we use individual cluster members as identified in data from Gaia DR3 \citep{GaiaCollaborationVallenari_2023} by \cite{Hunt24}. We search for X-ray counterparts from  \textit{Chandra} \citep{Weisskopf02}, XMM-Newton \citep{xmm}, eROSITA \citep{merloni24}, as well as radio counterparts from the Evolutionary Map of the Universe (EMU) survey \citep{2025PASA...42...71H} and the Karl G. Jansky Very Large Array \citep{2011ApJ...739L...1P}. We also take advantage of the Gaia variability flags to further classify sources, and we search for evidence of optical variability and further classification thanks to work by \cite{2023A&A...674A..13E}.

\subsection{Cluster sample}

The cluster sample is selected as follows; we targeted nearby OCs (within 1 kpc) observed in X-ray, targeting three different age ranges; a young OC, IC 2602 (30 Myr), a middle age OC, NGC 2632 (350 Myr), and an old cluster, NGC 2682/M67 (4 Gyr). IC 2602 and NGC 2632 are both within the publicly-released German eROSITA footprint, and M67 has been well studied in X-ray with ROSAT, \textit{Chandra} and XMM-Newton \citep{belloni98, vdb04, mooley15}. M67 is one of the most well studied open clusters in almost every wavelength (except radio), with optical studies to determine stellar cluster membership, and ultraviolet studies to find evidence of a blue straggler population \citep{1984ApJ...279..237P}.

Blue stragglers are bluer and brighter than the main sequence turnoff stars \citep{Sandage1953}. Stellar merger, collision, and mass-transfer are known to be the primary channels for their formation \citep{Boffin2015}.  Due to the low density of open clusters, stellar merger and mass-transfer are the only viable channels. The formation of blue stragglers due to multiple stellar interactions places them among the massive populations in star clusters \citep{shara1997}, and therefore they are representatives of dynamical ages of their host clusters as well \citep{Ferraro2012,rao2023}

We revisit M67 in light of new Gaia memberships. For our multi-wavelength study of M67, we can take advantage of \cite{2015AJ....150...97G}'s spectroscopic study of M67 members to search for radial velocity variability.   We report the cluster properties in Table 1.
\begin{table*}
\label{table:clusters}
\caption{Observed cluster properties from \cite{Hunt24}, except for the age of M67, as the ages are underestimated for clusters with blue straggler stars \citep{Hunt24, 2024AJ....167...12C}. M67 is estimated to be around 4 Gyr old \citep[][and references therein]{Reyes2024}.} 
\begin{tabular}{lllllll}
Cluster Name & RA           & Dec      &Age (Gyr)    & Dist (pc) & $r_J$ (deg) & Mass (\(M_\odot\))   \\ \hline
IC 2602      & 160.97308272 & $-$64.39323626 & 0.03&150.6     & 3.24     & 344  $\pm$ 42  \\
NGC 2632     & 130.08810396 & +19.66587324 & 0.75&183.5     & 4.28     & 1012 $\pm$ 74  \\
NGC 2682/M67 & 132.84984192 & +11.81745461 & 4&837.3     & 1.41     & 2761 $\pm$403 
\end{tabular}
\end{table*}

\begin{table*}
   
    \begin{tabular}{lll}
    Gaia ID & class name & period\\
      5237279792173016832& RS& - \\
      5239251727594949120 & SOLAR\_LIKE& - \\
      5239525196750143616&RS&-\\
      
      %5239626420542800512 & RS& - \\
      5239701569617071872 & YSO&-\\
      5239758155778687360 & RS &-\\
      5239851962193200896 & SOLAR\_LIKE & - \\
      5241082315708039168 & YSO & - \\
      
      5241621385657305856 & YSO &-\\
      5239242660940082432 & COMP\_VAR\_VSX\_2019& 1.19 \\
       %5239851962193200896 & COMP\_VAR\_VSX\_2019& - \\
       5239723525484923904& COMP\_VAR\_VSX\_2019& -\\
       5251802519709841280&COMP\_VAR\_VSX\_2019& - \\
       5239746271633131904& ASASSN\_VAR\_JAYASINGHE\_2019&2.00\\
       5242010093028921856& COMP\_VAR\_VSX\_2019& -\\
       5251805715165696384 &ASASSN\_VAR\_JAYASINGHE\_2019& 0.21 \\
       5239758155778687360 & COMP\_VAR\_VSX\_2019& - \\
       5241350287303624832 &COMP\_VAR\_VSX\_2019 & -\\
       5239498744077038976 &ASASSN\_VAR\_JAYASINGHE\_2019& 0.62 \\
       %5239701569617071872& COMP\_VAR\_VSX\_2019& - \\
       5244883529229180928&ASASSN\_VAR\_JAYASINGHE\_2019& 1.83 \\
       5239401432972879872&COMP\_VAR\_VSX\_2019& 5.93 \\
       5239626420542800512&ASASSN\_VAR\_JAYASINGHE\_2019 (RS)& 0.99
    \end{tabular}
    \caption{Variability flags and optical periods for X-ray sources in IC 2602, where reported in \cite{2023A&A...674A..13E}. Less than half of these X-ray sources have detected variability.}
    \label{tab:ic2602variability}
\end{table*}

\begin{table*}

    \begin{tabular}{lll}
    Gaia ID & class name & period \\
       661190015991474688, & ECL & - \\
       660225916090501376 &GAIA\_ROT\_GAIA\_2017 &5.85  \\
       661190153430426880 & RS&9.76 \\
       664453778817920256 &ASASSN\_VAR\_JAYASINGHE\_2019& - \\
       661458365546507776 &SOLAR\_LIKE&8.90\\
       661311752544249088, &GAIA\_ROT\_GAIA\_2017& 2.99 \\
       661408818803811200 &SOLAR\_LIKE&12.22 \\
       661216743570426240 &GAIA\_ROT\_GAIA\_2017&3.23 \\
       664288577196528128&SB9\_SB\_POURBAIX\_2004& 5.97\\
       661268940310277888&GAIA\_ROT\_GAIA\_2017&4.14 \\
       661289105178505344 &SB9\_SB\_POURBAIX\_2004& 45.98 \\
       661271173693364864&COMP\_VAR\_VSX\_2019& - \\
       661306319407420160 &GAIA\_ROT\_GAIA\_2017 &6.04\\
       663140171661012480 &KEPLERGAIA\_BY\_ROT\_DISTEFANO\_2023& 1.56\\
       659645030354241536 &GAIA\_ROT\_GAIA\_2017 & 11.29\\
       661294332158553856 & GAIA\_ROT\_GAIA\_2017 & 7.14\\
       664436770747504128 &GAIA\_ROT\_GAIA\_2017 & 12.78\\
       660998975844267264& SOLAR\_LIKE & 9.12 \\
       661419264165477504 & GAIA\_ROT\_GAIA\_2017 & 2.59 \\
       665129291274749696&GAIA\_ROT\_GAIA\_2017 &9.50 \\
       664324684984105728 &SB9\_SB\_POURBAIX\_2004 & 13.28\\
       660939258619177984 &GAIA\_ROT\_GAIA\_2017& 12.62 \\
       661148268907314432 &SOLAR\_LIKE& 4.53 \\
       661396754238802816  & HIP\_VAR\_ESA\_1997&  - \\
       664478620908806784 & SOLAR\_LIKE - \\
       664710686582129536 & SOLAR\_LIKE \\
    \end{tabular}
    \caption{Gaia variability flags and optical periods for X-ray sources in NGC 2632 (where available) from \cite{2023A&A...674A..13E}. The majority of the X-ray sources in NGC 2632 are from rotational variables. }
    \label{tab:ngc2632variability}
\end{table*}

\begin{table*}
\begin{tabular}{llll}
GAIA\_ID           & Mm & class name & period   \\
604906771677660544 & (BL)M & CATALINA\_VAR\_DRAKE\_2014 (ECL) & 0.51 \\
604911307163200000 & BN  &COMP\_VAR\_VSX\_2019 & - \\
604911509025877248 & (BL)M &ATLAS\_VAR\_HEINZE\_2018 (DSCT|GDOR|SXPHE) &0.36 \\
604914983655019520 & BM & LAMOST\_RAD\_VEL\_VAR\_TIAN\_2020 &  -  \\
604916422468464768 & BM  &SB9\_SB\_POURBAIX\_2004 & 4.36 \\
604916559907415040 & BM &SB9\_SB\_POURBAIX\_2004 & 31.78 \\
604917285757663872 & BM    \\
604917354477131392 & BM   & SOLAR\_LIKE \\
604917491916095872 & BM  &SB9\_SB\_POURBAIX\_2004(RS) & 10.06\\
604917526275831040 & BM    \\
604917629355039360 & BM  & COMP\_VAR\_VSX\_2019 & - \\
604917629355038848 & BM &SB9\_SB\_POURBAIX\_2004 & 7.16   \\
604917663714774784 & SM    \\
604917728138508160 & BM & SB9\_SB\_POURBAIX\_2004 &42.83   \\
604917934296938240 & BM   & SB9\_SB\_POURBAIX\_2004 & 11.02 \\
604918041671889792 & (BL)M &ASASSN\_VAR\_JAYASINGHE\_2019 (ECL)& 0.44\\
604918179110923392 & SM    \\
604918213470564992 & (BL)M &CATALINA\_VAR\_DRAKE\_2014 (ECL) &0.36 \\
604920683076006272 & BM &SB9\_SB\_POURBAIX\_2004 & 7.65   \\
604920756091231488 & (BL)M  & SOLAR\_LIKE \\
604920824810533632 & (BL)N  & RS \\
604921030968952832 & BM & SB9\_SB\_POURBAIX\_2004 (RS)& 18.39   \\
604921374566324992 & BM  &COMP\_VAR\_VSX\_2019 & 2.6  \\
604921374566321920 & BM  &SB9\_SB\_POURBAIX\_2004 & 1495.0  \\
604921855602675968 & BM  &COMP\_VAR\_VSX\_2019 & 1.44  \\
604923947251366656 & U &LINEAR\_VAR\_PALAVERSA\_2013 (RS) & 0.27  \\
604969787437702784 & U & SOLAR\_LIKE &0.26
	
\end{tabular}
\caption{Variability flags and optical periods compiled by \cite{2023A&A...674A..13E} along with binary membership from \cite{2015AJ....150...97G} for M67 X-ray sources.  (BL)M stands for likely binary member, BM for binary member, SM for single member,  (BL)N for \textbf{binary likely non-members, and U for unknown.} }
\end{table*}
\subsection{X-ray}
Due to their extended size (often spanning several degrees), nearby OCs are often not the target of detailed X-ray studies (with the exception of M67, which has been well studied at the center), but benefit from sensitive all-sky X-ray surveys like eROSITA, which can provide a consistent picture of the entire cluster. For IC 2602 and NGC 2632, we searched for X-ray counterparts to individual cluster members in eROSITA. We rejected three matches in IC 2602 because the optical magnitude was brighter than 6, and the X-rays were likely due to optical loading \citep[for further discussion, see][]{merloni24,2022A&A...661A..35S}.

M67 is nominally in the publicly available eROSITA footprint. However, we did not identify any X-ray counterparts from eROSITA. We therefore made use of archival pointed observations from \textit{Chandra} and \textit{XMM-Newton}. For \textit{Chandra}, we used the \textit{Chandra} Source Catalog version~2.1 \citep{Evans10}, which contains two relevant observations: ObsID~1873 (50~ks, 2001 May 31, PI: Belloni) and ObsID~17020 (10~ks, 2015 February 26, PI: van den Berg). For \textit{XMM-Newton}, we examined the European Photon Imaging Camera (EPIC) Pipeline Processed Source (PPS) catalogs for ObsID~0212080601 (15~ks, 2005 May 8, PI: Jansen) and ObsID~0109461001 (10~ks, 2001 October 20, PI: Mason). 

%and we used the Chandra Source Catalog version 2.1 \citep{Evans10} which contains both Chandra observations of M67 (ObsID 1873, 50ks, 2001-05-31, PI: Belloni, and ObsID 17020, 10ks, 2015-02-26, PI: van den Berg). We also searched the XMM-Newton archive and used the European Photon Imaging Camera (EPIC) pipeline processed (PPS) catalog of ObsID 0212080601 (15ks, 2005-05-08, PI: Jansen) and ObsID 0109461001 (10ks, 2001-10-20, PI: Mason).

%\textcolor{red}{K -- can you add a paragraph here about extracting the M67 source}
We were particularly interested in follow up on a specific X-ray source in M67, WOCS 3012/S1077. 
We extracted it from the merged \textit{Chandra} dataset and estimated its unabsorbed fluxes using the \texttt{srcflux} script in \texttt{CIAO}. A power-law spectral model with a fixed photon index of $\Gamma = 2$ was adopted, and the hydrogen column density was fixed at $N_H = 2.2\times10^{20}$~cm$^{-2}$ \citep{vdb04}.

For ObsID~1873 (effective exposure 46.26~ks), the unabsorbed flux in the 0.5–8.0~keV band was $7.8^{+0.49}_{-0.49}\times10^{-14}$~erg~cm$^{-2}$~s$^{-1}$. For the same model, the fluxes in the 1–10~keV and 0.5–10~keV bands were $4.75^{+0.40}_{-0.40}\times10^{-14}$ and $7.87^{+0.50}_{-0.49}\times10^{-14}$~erg~cm$^{-2}$~s$^{-1}$, respectively. Assuming a distance of 820~pc to M67 \citep{vdb04}, this flux corresponds to an X-ray luminosity of $L_X\approx6.3 \times 10^{30}$~erg~s$^{-1}$ in the energy band of 0.5--8.0~keV. 

In contrast, no significant X-ray emission was detected at the same coordinates in ObsID~17020 (effective exposure 9.82~ks). Using the Bayesian prescription for Poisson statistics in the case of zero source counts \citep{Kraft_1991}, we derived a 90\% confidence count-rate upper limit of 
\(2.34\times10^{-4}~\mathrm{counts~s^{-1}}\). 
Assuming the same spectral model as above (\(\Gamma = 2\), \(N_{H} = 2.2\times 10^{20}~\mathrm{cm^{-2}}\)), the corresponding unabsorbed flux upper limits are 
\(9.4\times 10^{-16}~\mathrm{erg~cm^{-2}~s^{-1}}\) (0.5--10~keV) and 
\(8.2\times 10^{-16}~\mathrm{erg~cm^{-2}~s^{-1}}\) (1--10~keV), which translate to luminosity limits of
\(L_X < 7.6\times 10^{28}~\mathrm{erg~s^{-1}}\) and 
\(L_X < 6.6\times 10^{28}~\mathrm{erg~s^{-1}}\), respectively. 

\subsection{Radio}
IC 2602 has been observed by the \textit{Australian SKA Pathfinder Telescope} EMU survey \citep{2025PASA...42...71H}. We searched for radio counterparts to X-ray point sources associated with cluster stars in the pipeline-processed catalogs. The EMU survey will mainly be sensitive to background AGN, but at the 150 pc distance of IC 2602, it is also sensitive to radio emission produced by chromospherically active contact binaries.

Of the 77 X-ray sources with high probability matches to cluster stars, only three had detected radio counterparts in EMU, all in the central region of IC 2602.%, in the pointing at the center of the cluster EMU 1050-64.SB54771. 
\begin{table*}[]
\caption{IC 2602 cluster members with X-ray from eROSITA and radio from the EMU survey. }
\begin{tabular}{llll}
GAIA\_ID       & G\_mag      & X-ray Flux & Radio Flux Density  \\
&& erg/s (0.2-2.3 keV) &mJy (943.5 MHz)\\  \hline
5253546997989686912 & 10.32 & 6.7 $\times 10^{-13}$       $\pm$ 4.7 $\times 10^{-14}$      & 0.18      $\pm$ 0.08          \\
5251888522138694656 & 11.25 & 5.0 $\times 10^{-13}$     $\pm$ 4.1  $\times 10^{-14} $      & 0.21     $\pm$ 0.02            \\
5239674700295136000 & 12.23  & 1.6 $\times 10^{-13}$      $\pm$ 2.2 $\times 10^{-14}$          & 0.14   $\pm$ 0.03          
\end{tabular}
\end{table*}

NGC 2632 and M67 are
too far north to be observed by ASKAP. However, M67 has been observed by VLA on December 11, 2010 (Project code: VLA/10B-173) for four hours in total. The observations were taken with the C-band receiver (centered at 5 GHz) with a 2 GHz total bandwidth. 3C286 (J1331+3030) was observed and used for bandpass and flux scale calibration and J0842+1835 for
gain calibration. The data were calibrated and imaged with CASA \citep{CASA2022}. The VLA Pipeline 2024.1.1 (CASA 6.6.1)\footnote{\url{https://science.nrao.edu/facilities/vla/data-processing/pipeline}} was used to perform flagging and calibration. Briggs weighting with robustness parameter 0 was used to image the data.  
%\textcolor{orange}{detail about the VLA data from Lilia, VLA/10B-173 , Dec 12,2010. 06:50-10:50 )(4 hours total, J1331+3030, phase cal J0842+1835 C band centered at 5GHz with ~2GHz bandwidth, Calibrated and imaged with casa. flagging and calibration performed using the VLA casa pipeline, robust weighting, robustness parameter} 

We used pyBDSF \citep{2015ascl.soft02007M}, to detect radio point sources down to a $\sim 5 \sigma$ threshold, and detected 50 unique point sources. We crossmatched VLA and Gaia data for potential matches within $1^{\prime\prime}$, and found four radio sources with a Gaia counterpart. Only one of these sources has cluster membership in M67 (604921855602675968, with a flux density of 96 $\pm 5 \mu$Jy). One is classified as a background AGN SDSS J085150.31+114855.9 (Gaia DR3 604918380973692800). Two have stellar classifications, Gaia DR3 604918174820102400 is classified as rotating variable EY Cnc, Gaia DR3 604915808290404480 is classified as a cluster member by \cite{1996AJ....112..628F}, but the X-ray counterpart from \cite{vdb04} is classified as a QSO, highlighting the prevalence of background AGN contamination in pre-Gaia cluster catalogs. 

\subsection{Crossmatching with \textsc{nway}}

We use \textsc{nway}\footnote{\url{https://github.com/JohannesBuchner/nway/}} \citep{2018MNRAS.473.4937S} to crossmatch between X-ray data from eROSITA and \textit{Chandra} and Gaia fields of IC 2602 and NGC 2632 (querying the full Gaia catalog from the position and Jacobi radius given in \citealt{Hunt24}). 

\textsc{nway} computes the match probability, useful in the case of multiple potential matches, and rules out sources with a high probability of chance superposition. Because these open clusters are not very crowded fields, we did not find any issues with crowding or multiple optical counterparts to an X-ray counterpart. \textsc{nway}'s match probability for all matches was very high ($>$ 95\% in all cases).

We crossmatched up to a radius of $5^{\prime\prime}$ for the following reasons: while the positional errors of eROSITA sources typically ranged from $1^{\prime\prime}$ up to $15^{\prime\prime}$, we do not report matches with separations larger than $5^{\prime\prime}$. We shifted the RA and Dec of one catalog in several directions by $10^{\prime\prime}$ and, after rematching, found that the shifted catalogs gave a large number of matches beyond $5^{\prime\prime}$, suggesting that matches with a separation higher than $5^{\prime\prime}$were contaminated.

For M67, thanks to \textit{Chandra}'s subarcsecond spatial resolution, we considered matches between Gaia and the CSC up to $1^{\prime\prime}$. We only found two XMM-Newton matches to M67 stellar members that were not also detected by \textit{Chandra}. We note that we expect to have different results to \cite{belloni98, vdb04} and \cite{mooley15}, as \cite{Hunt24}'s Gaia cluster catalogs are more accurate than those used in previous X-ray studies of M67.

\subsection{Optical Variability}
We searched for evidence of optical variability using the Gaia variability flags, and \cite{2023A&A...674A..13E} which crossmatches Gaia data to all known optical surveys. For each cluster, we search the optical counterpart to a given X-ray source in these databases, and we catalog their classification and period if one exists in the literature. A preview of these tables can be seen in the Appendix.

Gaia classifications include solar-like variability, eclipsing binaries, RS Canum Venaticorum, $\alpha^2$ CVn and associated stars, $\delta$ Scuti/$\gamma$ Doradus/SX Phoenicis, young stellar objects, RR Lyrae and slowly pulsating B stars. Thanks to longterm optical surveys such as \cite{2014ApJS..213....9D, vsx, heinze2018, Tian2020, pourbaix04, jayasinghe2019, palaversa13, esa1997} in some cases, we are able to obtain optical periods for the systems. For the X-ray sources, we report variability class and period (if known) in Tables 2,3, and 4.   

Of the 33 X-ray sources we found associated with a M67 individual star, 27 were found in the \cite{2015AJ....150...97G}     spectroscopic study. \textbf{Of these, 80\%} fall in the binary member category, or likely binary member category. Three are likely single members, and two are unknown (but still proper motion members). We report the \cite{2015AJ....150...97G} binary member category in Table 4, along with the variability classification from Gaia for M67.

\section{Results and Discussion} \label{sec:results}
We used new and archival multiwavelength observations to search the contents of three open clusters: IC 2602 (0.03 Gyr), NGC 2632 (0.35 Gyr) and M67 (4 Gyr).  
\begin{table*}

\caption{Gaia variability flags \citep{2023A&A...674A..13E} for stellar members of the three clusters: solar-like variability (SOLAR\_LIKE), eclipsing binaries (ECL), RS Canum Venaticorum (RS), $\alpha^2$ CVn and associated stars (ACV), $\delta$ Scuti/ $\gamma$ Doradus/SX Phoenicis stars (DSCT), young stellar objects (YSO), RR Lyrae (RR), and slowly pulsating B star (SBP). The main types of variability are either due to rotation (ACV, RS, SOLAR\_LIKE) or pulsation (DSCT, RR, SPB).}
\begin{tabular}{lllllllll}
Cluster Name & SOLAR$\_$LIKE & ECL & RS & ACV & DSCT & YSO & RR & SBP \\ \hline
IC 2602      & 5            & -   & 11 & -   & -    & 32  & 1  & 3   \\
NGC 2632     & 155          & 1   & 1  & 2   & 1    & -   & -  & -   \\
NGC 2682/M67 & 21           & 3   & 8  & 0   & 1    & -   & -  & -  
\end{tabular}
\end{table*}
\subsection{IC 2602}
\begin{figure}
 
    \includegraphics[width=3in]{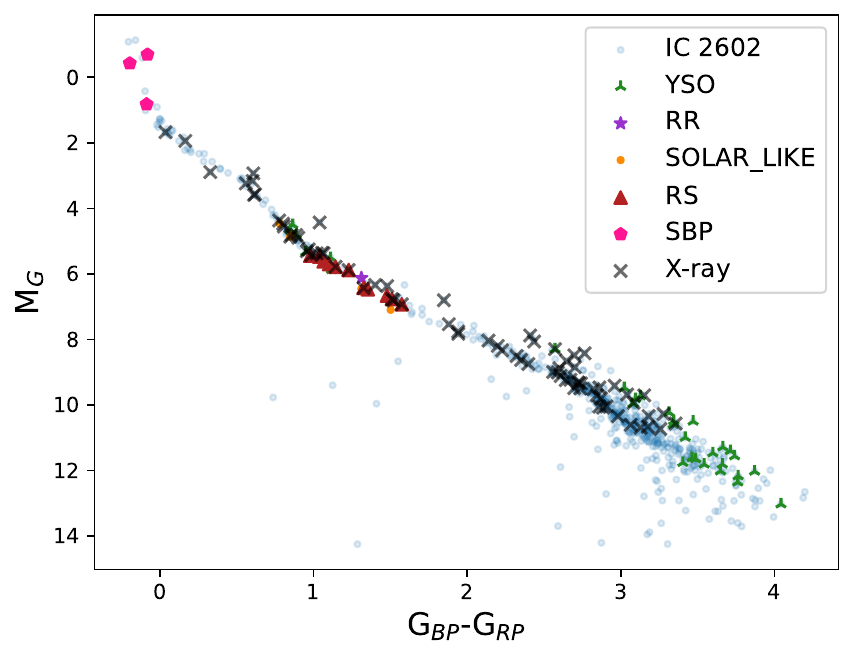}
    \caption{CMD for IC 2602. Black x's are X-rays from eROSITA, green triangles are $\delta$ Scuti/$\gamma$ Doradus/SX Pheonicis, purple pentagons are ACV systems, red triangles are RS Canum Venaticorum, teal squares are eclipsing binaries and orange points are solar-like variability. }
    \label{fig:ic2602cmd}
\end{figure}

\begin{figure}
   
    \includegraphics[width=4in]{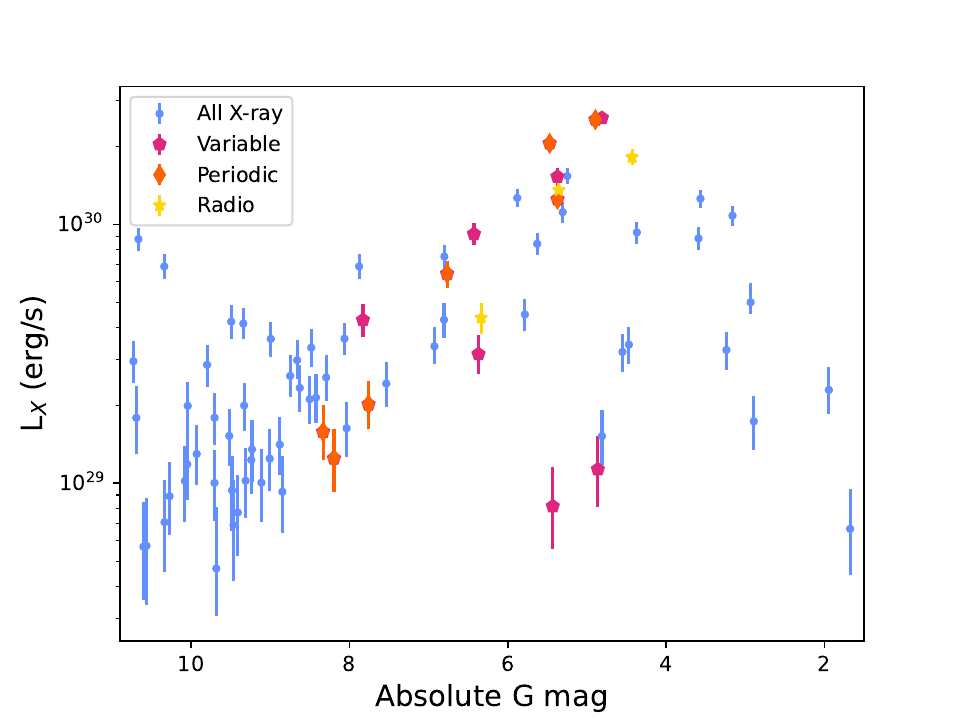}
    \caption{X-ray luminosity (eROSITA band) versus absolute G magnitude for X-ray sources in IC 2602. Sources with variability flags from \cite{2023A&A...674A..13E} are labeled in orange pentagons, with periodic sources marked with pink diamonds. Three sources (yellow triangles) had radio emission associated with them, but were not flagged as variable by \cite{2023A&A...674A..13E}. }
    \label{fig:ic2602-var}.
\end{figure}

Of the 77 X-ray sources associated with IC 2602, only 14 had variability classification or measured periods. Three sources had radio from EMU associated with them, and the radio and X-ray fluxes suggest these systems are chromospherically active binary stars, consistent with \citealt{1995A&A...302..775G,2024PASA...41...84D} (see also \citealt{Paduano24}). The radio and X-ray fluxes of these sources are summarized in Table 5. Other classification of the X-ray sources include young stellar objects, RS Canum Venaticorum, and solar-like sources. Several other of the X-ray sources are unclassified variable sources found in AAVSO and ASAS-SN. The location of these sources on the color-magnitude diagram (CMD) can be seen in Figure \ref{fig:ic2602cmd}, and the X-ray luminosity versus optical magnitude in Figure \ref{fig:ic2602-var}. The radio detections appear to be early to mid-F spectral type stars. Based on the CMD of IC 2602, these stars have masses between 1.2 and 1.5 solar masses, which indicates that they have begun to develop convective envelopes, signifying that their stellar structure is changing.

An upper main-sequence star with Gaia ID = 5239843200460432384 is identified as a very fast rotator and has also been classified as an SPB variable. This matched to a detection in eROSITA, which we rejected as contaminated as it fell into the optical loading range.  If this source is indeed X-ray bright, the combination of SPB variability, rapid rotation, and strong X-ray emission would point toward either the presence of a companion or a Be/Bp spectral classification. Be stars are a subset of B-type stars that host decretion disks driven by their rapid rotation, and a fraction of them are known to exhibit SPB-type pulsations \citep{Rivinius2016,Shi2023}. Additionally, two other stars, Gaia IDs = 5299121205191777792 and 5239841340702856960, show high RUWE values, strongly suggesting the presence of gravitational companions. Notably, Gaia ID = 5299121205191777792 is even listed in SIMBAD as a double or multiple system.

\subsection{NGC 2632}
\begin{figure*}
    
    \includegraphics[width=6in]{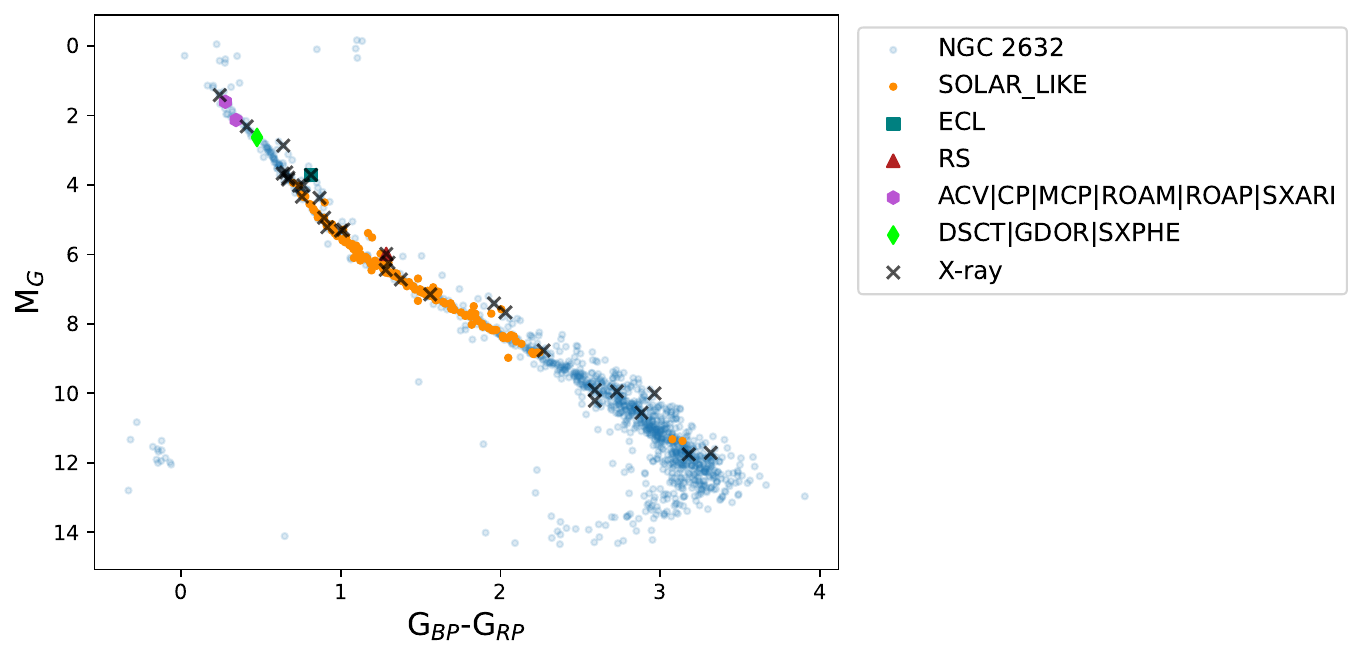}
    \caption{CMD for NGC 2632. Black x's are X-rays from eROSITA, green diamonds are $\delta$ Scuti/$\gamma$ Doradus/SX Pheonicis, purple pentagons are ACV stars, red triangles are RS Canum Venaticorum, teal squares are eclipsing binaries, and orange points are solar-like variability. }
    \label{fig:ngc2632cmd}
\end{figure*}

\begin{figure}
  
    \includegraphics[width=3.5in]{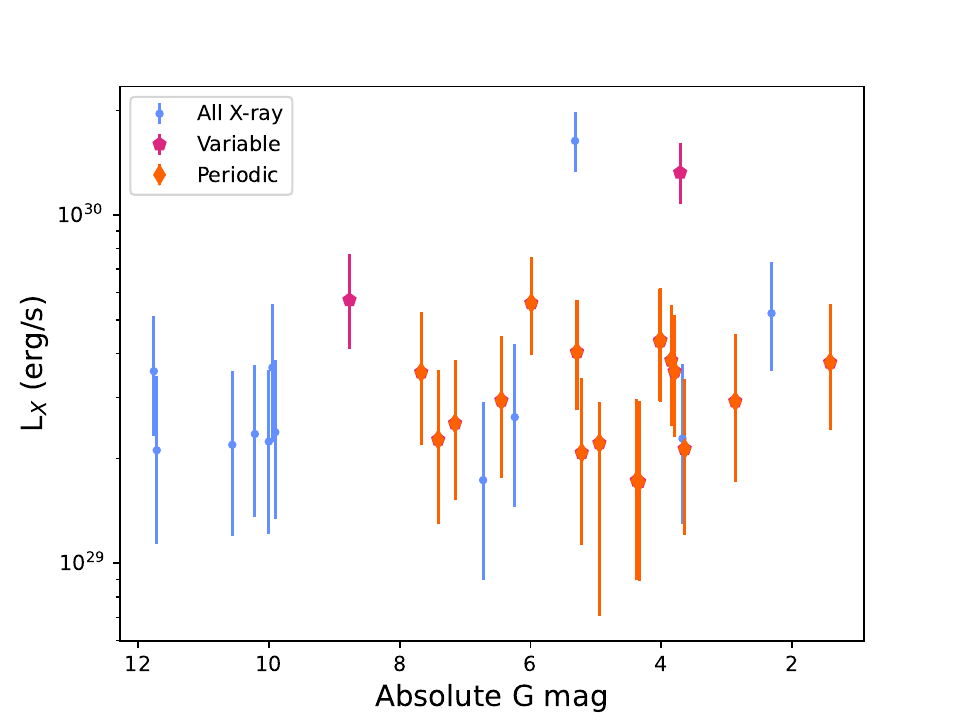}
    \caption{X-ray luminosity (0.5-2.3 keV) versus absolute G magnitude for X-ray sources in NGC 2632. Variable sources are denoted with orange pentagons, and periodic variable sources are marked with pink diamonds. }
    \label{fig:ngcc2632xray}
\end{figure}

We identified 31 eROSITA sources which can be associated with NGC 2632 stellar members. Of these, \textbf{19} showed optical variability. The majority of the X-ray systems have solar-like variability, with one eclipsing binary, one RS Canum Venaticorum, rotational sources, and other variability \citep{pourbaix04, jayasinghe2019, esa1997, 2023A&A...674A..20D}. The location of the X-ray sources and optical variables on the CMD can be seen in Figure \ref{fig:ngc2632cmd}, and we plot the X-ray luminosity versus optical magnitude in Figure \ref{fig:ngcc2632xray}.
We searched the VLASS all sky catalog \citep{2020PASP..132c5001L}, but there were no detected radio counterparts matching to eROSITA cluster associated X-ray sources within 10". 

One of the upper main-sequence X-ray sources in NGC 2632 (Gaia ID = 664314759317023360, $G = 8.632$ mag, corresponding to an absolute G mag of 2.31) has a $v_{\text{broad}} = 195.29 \pm 2.43$ km s$^{-1}$ and a high RUWE value of 5.12, suggesting the presence of a gravitationally bound companion. The observed X-ray emission may either arise from interaction processes within the system that is also driving rapid stellar rotation, or from a low-mass companion.

\subsection{M67}

\begin{figure*}
  
    \includegraphics[width=6in]{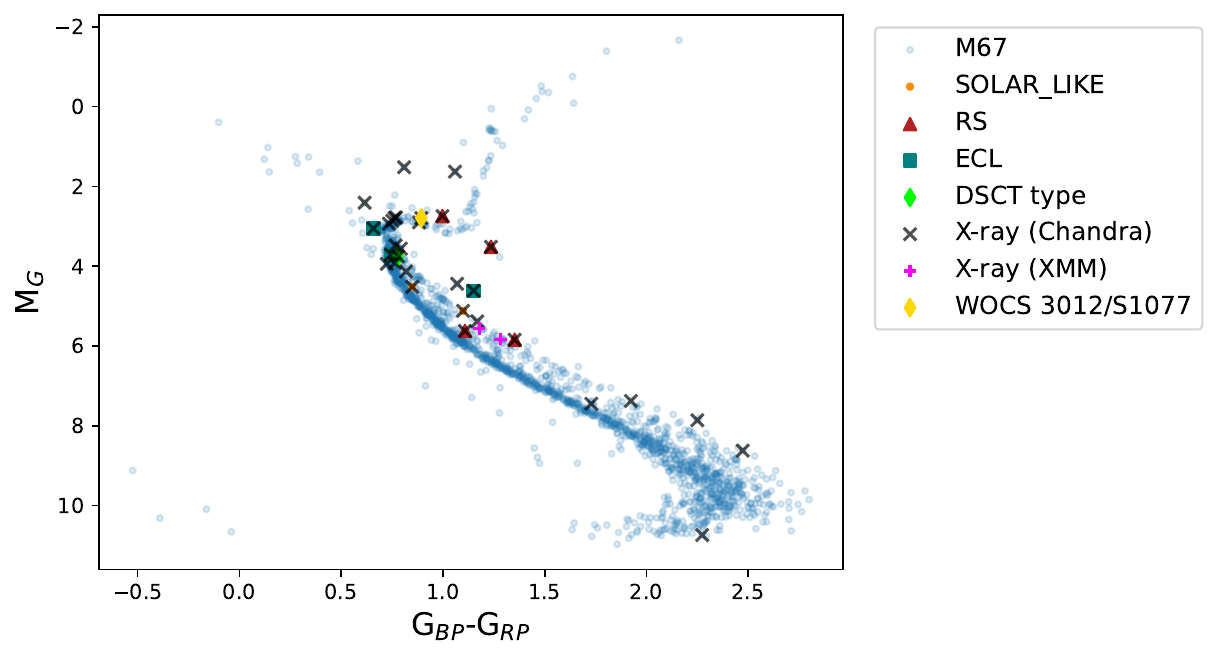}
    \caption{CMD for M67. X-rays from \textit{Chandra} are denoted in x's, and X-rays from XMM-Newton are denoted by purple plusses. Different Gaia variability classifications are shown, a green diamond for pulsating systems $\delta$ Scuti/$\gamma$ Doradus/SX Pheonicis, teal squares for eclipsing binaries (which are also X-ray sources), red triangles for RS Canum Venaticorum (which are also X-ray sources), and orange dots for solar-like variability. }
    \label{fig:m67-cmd}
\end{figure*}

\begin{figure}
  
    \includegraphics[width=3.5in]{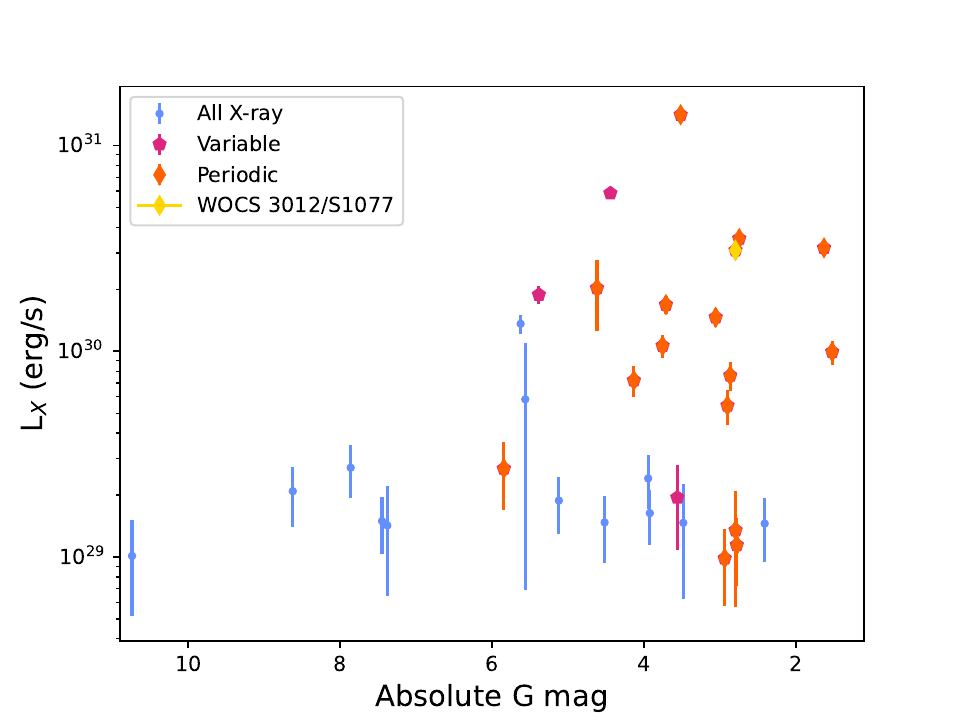}
    \caption{X-ray luminosity (0.5-7.0 keV) versus absolute G magnitude for M67's X-ray sources. Pink diamonds denote periodic variability, while orange pentagons show variability flags from \cite{2023A&A...674A..13E}. }
    \label{fig:m67xrat}
\end{figure}

%We crossmatched cluster associations from Hunt \& Reffert to X-rays from the Chandra Source Catalog (Evans..) and to XMM-Newton (Mooley 2015). 
In M67, we found 31 \textit{Chandra} sources associated with an individual cluster member; we plot these sources and the optical variables on M67's CMD in Figure \ref{fig:m67-cmd}, and the X-ray luminosity vs optical magnitude in Figure \ref{fig:m67xrat}. 

\begin{figure*}
  
    \includegraphics[width=7in]{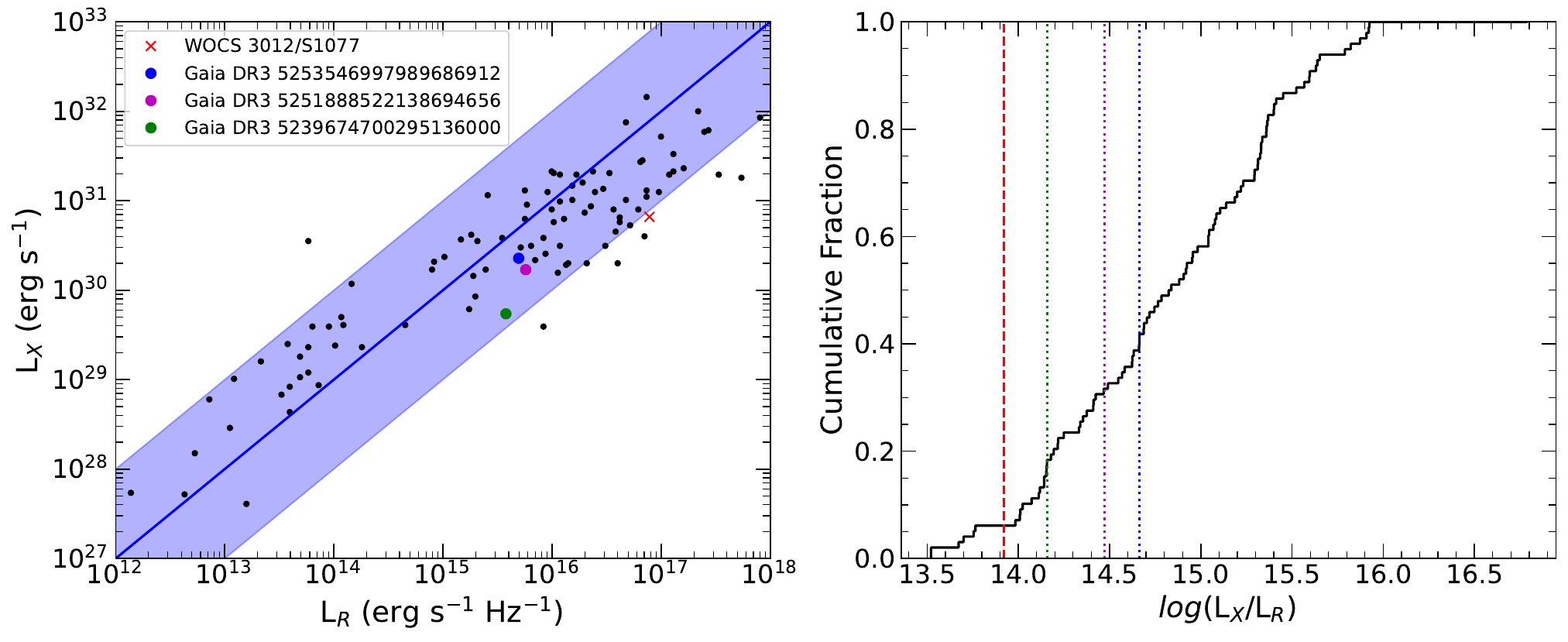}
    \caption{The location of IC 2602 radio/X-ray sources (Table 5) and WOCS 3012/S1077 on the relation of radio and X-ray for active binaries from \cite{1995A&A...302..775G}. Adapted from \cite{Paduano24}. The three sources in IC 2602 fall firmly on the correlation for active binaries. WOCS 3012/S1077 (dashed line in left hand side) falls in the scatter near the correlation. }
    \label{fig:gudelbenz}
\end{figure*}

\begin{figure*}
  
    \includegraphics[width=5in]{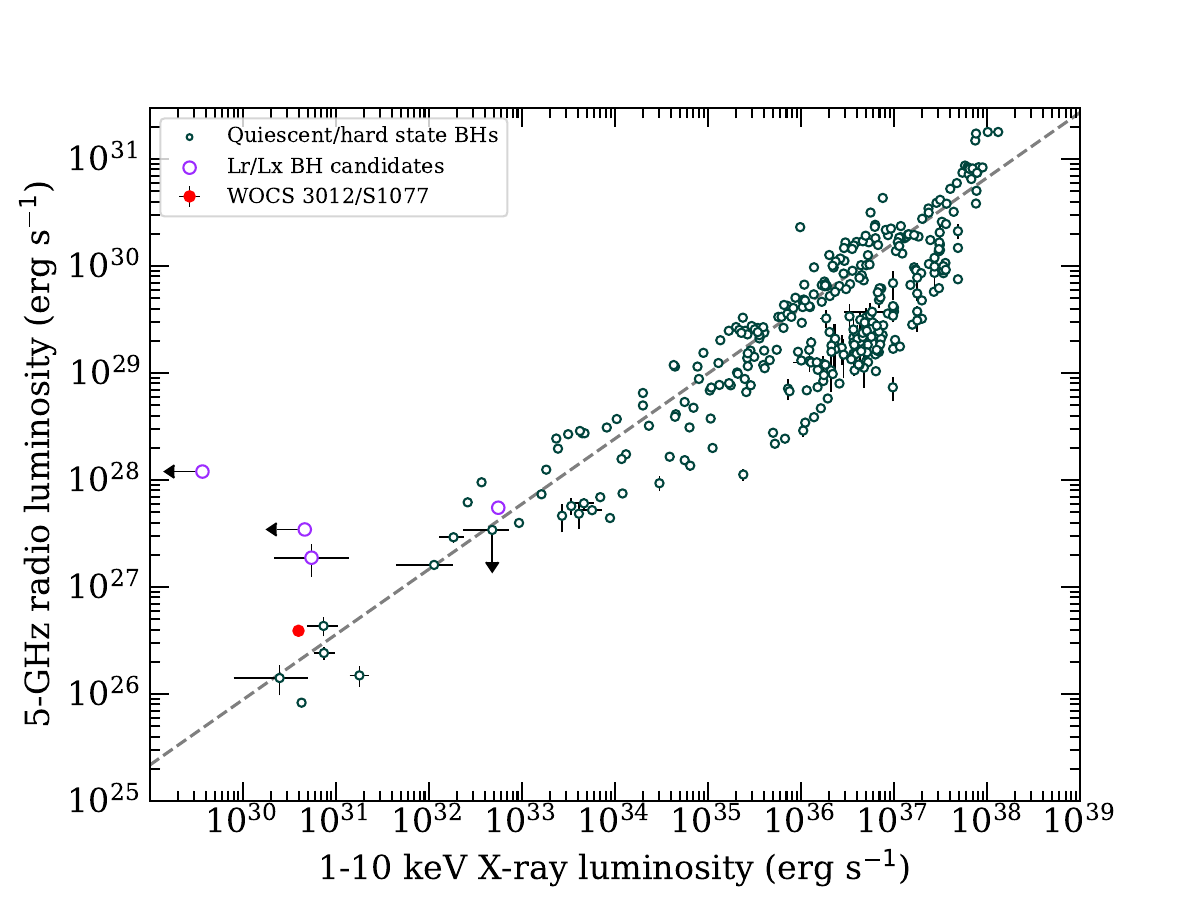}
    \caption{Location of WOCS 3012/S1077 on the radio/X-ray correlation for black holes, next to known quiescent black holes, and black hole candidates. WOCS 3012/S1077 occupies a space on this correlation that is near where one would expect to find a quiescent black hole, based on the X-ray and radio. In comparison, the three X-ray/radio sources identified in IC 2602 are much fainter in X-ray, and several orders of magnitude louder in radio than than known X-ray binaries, and are not near the correlation plane at all. Figure modified from \cite{Bahramian18}.}
    \label{fig:lrlx}
\end{figure*}
We note that our X-ray associations to M67 are significantly different from previous studies \citep{vdb04, mooley15}, because despite using the same X-ray observations, our cluster catalog is significantly different. In particular, \cite{belloni98}'s ROSAT study, which uses a very different ground-based cluster catalog (where the typical uncertainties are on the order of Gaia at 21st magnitude ($\sim$1 mas / yr), which is the limiting mag of Gaia). We do not attempt to replicate this particular study because the ROSAT positional uncertainty is significant (ranging from 3" to 29"), and although M67 as an open cluster is less crowded than a globular cluster, that positional uncertainty is still large enough that there can be multiple optical counterparts to a given X-ray source. Using \cite{belloni98}'s positions and uncertainties, we found that the only ROSAT sources with relatively small positional uncertainties, which crossmatched to an individual Gaia cluster member, were also found with the \textit{Chandra} data.

In M67, two X-ray sources that are brighter and bluer than the main sequence turnoff are blue stragglers, two X-ray sources located above sub-giant branch stars are evolved BSS/yellow stragglers, and one X-ray source located below the sub-giant branch is a sub-subgiant star (GaiaDR3 ID 604921030968952832). \citet{Geller2017} found two M67 sources as sub-subgiant stars, GaiaDR3 ID 604921030968952832 (S1063), GaiaDR3 ID 604972089540120832 (S1113), and both of those show X-ray emission and are binaries. 

Unlike IC 2602 and NGC 2632, where the X-ray studies are complete (if not contaminated due to eROSITA positional uncertainties), the \textit{Chandra} and XMM pointings only targeted near the center of the cluster, and thus our knowledge of M67's contents is less complete.

For the X-ray sources associated with  \cite{Hunt24}'s M67 catalog, 27 have been studied spectroscopically by \cite{2015AJ....150...97G}. 21 of these have binary membership classification. Many of these systems are variable, with Gaia variability flags solar-like, ECL, DSCT, RS, and have periods reported in the literature. We note that only 6 of the M67 X-ray sources do not have binary membership (Table 4).

From the VLA observation pointed at M67's center, we identified 50 radio point sources. Of these, only four matched to X-ray and optical sources detected by \textit{Chandra} and the complete Gaia catalog of the central region of M67. Two were classified as QSO/AGN, and were previously classified as M67 candidate members by \cite{1996AJ....112..628F}. Gaia DR3 604918174820102400 is not classified as a cluster member by \cite{Hunt24}, and is too faint to have parallax and proper motions measured, but has an 8.4 day optical period. This optical period may be more consistent with a non-member star than an AGN. 

Gaia DR3 604921855602675968 (WOCS 3012/S1077) is a cluster member, both from Gaia proper motions \citep{Hunt24}, but also confirmed via spectroscopy \citep{2015AJ....150...97G}. The 10GHz radio luminosity of this source, assuming the distance to M67, is $3.96 \pm 0.25 \times 10^{26}$ erg/s. We analyzed the radio and X-ray luminosities, comparing them to both chromospherically active binary stars \citep{1995A&A...302..775G}, (Figure \ref{fig:gudelbenz}) as well as quiescent black hole binaries in Figure \ref{fig:lrlx}. \textbf{In Figure \ref{fig:gudelbenz}, we } see that WOCS 3012/S1077 falls in the scatter of the luminous end of the \cite{1995A&A...302..775G} relationship for chromospherically active binaries\textbf{ (and on the flatter part of the cumulative distribution function)}, and also fits neatly on the radio/X-ray correlation for a quiescent X-ray binary. In contrast, the three radio/X-ray sources identified in IC 2602 do not fit on the X-ray binary correlation plane at all, but do fit in the \cite{1995A&A...302..775G} correlation.  

If the radio/X-ray emission from WOCS 3012/S1077 is due to a compact object, and is not of stellar origin, it is more likely to be from a black hole than a neutron star. While neutron stars do sometimes occupy the location on the radio/X-ray correlation generally occupied by black holes, very few neutron stars produce X-ray luminosity this faint \citep{Heinke2003, Heinke2006, Bahramian2015, degenaar17,vde21, postnov22}. Because the VLA observation of M67 was only a 4 hour exposure, we also cannot currently rule out that the radio emission is caused by a radio flare star \citep[e.g.,][]{2024PASA...41...84D}. 

%\cite{2021AJ....161..190G}'s spectroscopic study also reveals that WOCS 3012/S1077 is one of at least 6 spectroscopic triple systems uncovered in M67, with an inner orbit of 1.4 days, and an outer orbit of roughly 10 years. The inner orbit has a mass ratio of 0.76, and the inner visible star is at most 1.3 $M_\odot$ based on its position on the main sequence, suggesting that the unseen inner companion is $\sim$ 0.9 $M_\sun$. WOCS 3012/S1077 also shows a FUV excess \citep{2018MNRAS.481..226S}.

Like the putative black hole discovered by \cite{Paduano24}, WOCS 3012/S1077's radio and X-ray emission have ambiguous interpretations, where it falls into both chromospherically active binary stars, and quiescent stellar mass black holes. Spectroscopy by \cite{2021AJ....161..190G} provides a mass ratio of the inner binary (q=m2/m1) as 0.76. We assume, based on its location on the CMD, that the inner visible star has just left the main sequence, and the mass is between 1.2 and 1.3 $M_\odot$, suggesting that the unseen secondary has a minimum mass that is roughly solar, but the object's true mass is highly dependent on the inclination angle, which is not constrained. Therefore, this spectroscopic information does not exclude a less luminous star as the unseen companion to the inner binary, and also allows for a neutron star or black hole to explain the system, in the case of a very extreme inclination angle ($> 90$\%).

The long-term X-ray variability, dropping by around two orders of magnitude in X-ray over $\sim$ 15 years may provide a hint as to the nature of the unseen inner binary member. While chromospherically active binaries are known to exhibit X-ray variability \citep[e.g.,][among others]{1999ApJ...524..988K, 2018A&A...616A.161P}, a change of two orders of magnitude of X-ray luminosity is highly consistent with behaviors in the accretion processes of black holes and neutron stars \citep[e.g.,][of many notable examples]{2016ApJS..222...15T, Panurach21}. 

However, the Kepler K2 photometric lightcurve of this source (EPIC 211416111) exhibits optical variability profiles consistent with chromospherically active binaries, with clear evidence of significant stellar flaring activity. Several optical flares, including a superflare, were identified through visual inspection, all displaying the characteristic flare profile of a rapid rise followed by a relatively slower decay. In addition, notable variations in the lightcurve shape suggest varying starspot activity. Taken together, these findings indicate that this source is likely a chromospherically active binary, rather than a quiescent black hole.  However, if the chromosperic activity is coming from a very spun-up primary star at an extreme inclination angle, then it would be possible for this system to also host a quiescent stellar mass black hole.

\section{Summary and Conclusions} \label{sec:summary}
We take advantage of the wealth of multiwavelength surveys, from X-ray to optical to radio, and we search for multiwavelength contents of three open clusters, IC 2602 (30 Myr), NGC 2632 (0.75 Gyr) and M67 (4 Gyr), using archival X-ray and radio data, and \cite{Hunt24}'s Gaia catalog of cluster members.  We identified 77 X-ray sources in IC 2602, many of which were variable systems, and detected evidence of chromospherically active binaries in the EMU survey.  In NGC 2632, we found 31 X-ray sources, 27 of which were variable. 

NGC 2632 contained the largest number of variable stars (155 stars with solar-like variability). IC 2602 contains a large number of young solar objects, 1 RR Lyrae star, 3 slowly pulsating B stars, and 11 RS Canum Venaticorum. In contrast, M67 contains fewer variable systems, only 21 stars with solar-like variability, and 8 RS Canum Venaticorum stars, with 3 eclipsing binaries, and 1 pulsating DSCT. The X-ray contents of the clusters with known optical variability are presented in Tables 2, 3, and 4, and the overall summary of the optical variability is summarized in Table 6. 

The global study of optical variability in \cite{2025arXiv250812866A} shows that the fraction of types of variables is dependent on age; young systems like IC 2602 are expected to have a significant number of YSOs. Older systems like NGC 2632 will have a large fraction of solar-like variables, which we also see. At gigayear ages like M67, one expects a rather low fraction of RS systems, however, that is the second highest amount of variable systems seen in M67, highlighting its uniqueness as an old and massive open cluster. Gaia gives us a chance to do a population study of a massive amount of open clusters in the optical, and it is currently unknown how much radio and X-ray flux can be observed as a function of age. With the advent of all sky X-ray and radio surveys, this question can be addressed for the first time. With a larger sample, it may be possible to understand the detailed properties of a cluster's contents by its overall X-ray flux. 

M67 may harbor more radio and X-ray sources than the 31 cluster X-ray sources detected, but with the current coverage, we do not yet have a complete picture of its contents, but what we do know is already illuminating. 

Modeling dynamical stellar evolution in clusters is one of the most pressing topics in modern astronomy; knowing the behaviors of all individual cluster stellar members is vital for benchmarking simulations of dynamical evolution in star clusters \citep{Hurley02, Hurley07}. After around 10 Myr, black holes and neutron stars are expected to start forming in a star cluster system, but their final fates, and if they are retained in the clusters at all are unknown, because their natal kicks are currently not well constrained. Therefore, it is especially interesting to examine these clusters for potential black holes/neutron stars. 

This paper demonstrates that by combining modern X-ray and radio surveys, along with new optical cluster studies, it is possible to be sensitive to the high energy emission of compact objects in the very wide variety of open clusters in the Milky Way. Thanks to the wide variety of optical data available on these stars, it is also possible to probe a black hole origin for the high energy emission, as well as capture the broad range of stellar activity in these clusters, which all serve the purpose of better improving models of dynamical evolution in star clusters. 

\textbf{The authors thank the referee for their helpful comments which greatly improved the manuscript.} KCD thanks Aaron Geller for helpful discussion, and is indebted to Susmita Sett for her LaTeX skills.
This work has made use of data from the European Space Agency (ESA) mission {\it Gaia} (\url{https://www.cosmos.esa.int/gaia}), processed by the {\it Gaia} Data Processing and Analysis Consortium (DPAC,
\url{https://www.cosmos.esa.int/web/gaia/dpac/consortium}). Funding for the DPAC has been provided by national institutions, in particular the institutions participating in the {\it Gaia} Multilateral Agreement. This work is based on data from eROSITA, the soft X-ray instrument aboard SRG, a joint Russian-German science mission supported by the Russian Space Agency (Roskosmos), in the interests of the Russian Academy of Sciences represented by its Space Research Institute (IKI), and the Deutsches Zentrum für Luft- und Raumfahrt (DLR). The SRG spacecraft was built by Lavochkin Association (NPOL) and its subcontractors, and is operated by NPOL with support from the Max Planck Institute for Extraterrestrial Physics (MPE). The development and construction of the eROSITA X-ray instrument was led by MPE, with contributions from the Dr. Karl Remeis Observatory Bamberg \& ECAP (FAU Erlangen-Nuernberg), the University of Hamburg Observatory, the Leibniz Institute for Astrophysics Potsdam (AIP), and the Institute for Astronomy and Astrophysics of the University of Tübingen, with the support of DLR and the Max Planck Society. The Argelander Institute for Astronomy of the University of Bonn and the Ludwig Maximilians Universität Munich also participated in the science preparation for eROSITA.
Facilities: Gaia, XMM-Newton, eROSITA, \textit{Chandra} X-ray Observatory, ASKAP, VLA
Software: astropy \citep{Robitaille13}, CASA \citep{CASA2022}, matplotlib \citep{Hunter07}, NumPy \citep{harris20}, pandas \citep{Mckinney10}, LSDB \citep{2025arXiv250102103C}

%\facilities{Gaia, XMM-Newton, eROSITA, \textit{Chandra} X-ray Observatory, ASKAP, VLA}

%\software{astropy \citep{Robitaille13}, CASA \citep{CASA2022}, matplotlib \citep{Hunter07}, NumPy \citep{harris20}, pandas \citep{Mckinney10}, LSDB \citep{2025arXiv250102103C}}

\section{Appendix}
We include the crossmatch catalogs of X-ray and optical for IC 2602, NGC 2632 and M67.  The X-ray information for IC 2602 and NGC 2632 come from eROSITA, and the X-rays for M67 come from \textit{Chandra}. All three clusters were crossmatched with Gaia, and sub-selected for cluster membership based on \cite{Hunt24}. We include eRASS ID, R.A \& Dec and positional error from eROSITA, along with the 0.5-2.3 keV flux in erg/s and flux error, and the Gaia ID, R.A. and Dec from optical, along with the maximum match separation. For M67, we include the same information from Gaia and NWAY, and include the \textit{Chandra} positions and fluxes. A preview of these tables is displayed below.

\begin{sidewaystable*}[]
\begin{tabular}{llllllllllll}
Gaia DR3 ID         & R.A. (Gaia)        & Dec (Gaia)         & eRASS ID                & R.A. (eRASS)       & Dec  (eRASS)        & Err. & Flux          & Err.       & Sep.  \\ \hline
5249783644555618432 & 148.645965 & -62.900093 & 1eRASS J095435.0-625400 & 148.646056  & -62.900183 & 9.39  & 6.59E-14  & 1.98E-14  & 0.36    \\
5253742607985183872 & 158.189693 & -62.345237 & 1eRASS J103245.2-622044 & 158.188495 & -62.345570 & 6.10 & 3.76E-14 & 1.26E-14 & 2.33  \\
5238755920895671040 & 164.546205 & -65.713997 & 1eRASS J105811.1-654253 & 164.546500  & -65.714733  & 6.4 & 1.72E-14 & 9.26E-15   & 2.68  \\
5240531632175135616 & 168.747405  & -64.347075 & 1eRASS J111459.3-642049 & 168.747367 & -64.346954  & 3.25 & 2.84E-14  & 1.02E-14 & 0.44  \\
5239242660940082432 & 159.894951 & -66.135807 & 1eRASS J103934.7-660809 & 159.894713 & -66.135931  & 3.17 & 5.81E-14  & 1.42E-14   & 0.56 
\end{tabular}
\end{sidewaystable*}

\begin{sidewaystable*}[]
\begin{tabular}{llllllllll}
GAIA DR3          &  R.A. (Gaia)          &  Dec. (Gaia)     & \textit{Chandra} ID       &  R.A. (Chandra)      & Dec. (Chandra)    & Flux  & Sep. \\ \hline
604906771677660544 & 133.070113  & 11.808667 & 2CXO J085216.8+114831 & 133.069971 & 11.8087465  & 2.35 $\pm 0.9 $ E-14           & 0.57      \\
604911307163200000 & 132.823794 & 11.741530 & 2CXO J085117.7+114429 & 132.823832 & 11.741510 & 2.19 $\pm$ 0.20 E-14            & 0.15      \\
604911509025877248 & 132.770062 & 11.765787 & 2CXO J085104.8+114556 & 132.770070 & 11.7657855 & 1.23 $\pm$ 0.16 E-14             & 0.03     \\
604911680824692480 & 132.797490  & 11.786379 & 2CXO J085111.3+114710 & 132.797398 & 11.786334 & 3.16 $\pm$ 1.09 E-15           & 0.36         \\
604914983655019520 & 132.758790 & 11.817018& 2CXO J085102.0+114901 & 132.758592 & 11.817090 & 2.26 $\pm$ 1.0 E-15          & 0.75      
\end{tabular}
\end{sidewaystable*}

\paragraph{Data Availability Statement}
All data is publicly available. 

% A statement about how to access data, code and other materials allowing users to understand, verify and replicate findings --- e.g. Replication data and code can be found in Harvard Dataverse: \verb+\url{https://doi.org/link}+.

%\endnote in some journals will behave like \footnote; and \printendnotes will not output anything. 
\printendnotes

%\printbibliography
\bibliography{example}
% \appendix
% \section{Example Appendix Section}

\end{document}